\documentclass[aip,pop,amsmath,amssymb,preprint]{revtex4-1}
\usepackage[toc,page]{appendix}
\usepackage{lipsum}
\usepackage{subcaption}
\usepackage{hyperref}
\usepackage{graphicx}
\usepackage{dcolumn}
\usepackage{bm}
\usepackage{xcolor}
\usepackage{amsmath}
\usepackage[T1]{fontenc}
\usepackage{mathptmx}

\newcommand{\RN}[1]{%
  \textup{\uppercase\expandafter{\romannumeral#1}}%
}
\begin{document}

\author{Amirhassan Chatraee Azizabadi}
\email{amirchatraee@yahoo.com}
\affiliation{Department of Physics, Freie Universit\"at Berlin, Arnimallee. 14, 
14195 Berlin,
Germany}
\affiliation{Zentrum f\"ur Astronomie und Astrophysik, 
Technische Universit\"at Berlin, 
Hardenbergstr. 36, 
10623 Berlin, Germany}
\author{Neeraj Jain}
\affiliation{Zentrum f\"ur Astronomie und Astrophysik, 
Technische Universit\"at Berlin, 
Hardenbergstr. 36, 
10623 Berlin,
Germany}

\author{J\"org B\"uchner}
\affiliation{Zentrum f\"ur Astronomie und Astrophysik, Technische Universit\"at Berlin, 
Hardenbergstr. 36, 
10623 Berlin,
Germany}
\affiliation{Max-Planck-Institute for Solar System Research,
Justus-von-Liebig Weg 3, 
37077 G\"ottingen,
Germany}

\title{Identification and characterization of current sheets in collisionless plasma turbulence}
\date{\today}

\begin{abstract}
The properties of current sheets forming in a ion-kinetically turbulent 
collisionless plasma are investigated by utilizing the results of two-dimensional 
hybrid-kinetic numerical simulations.
For this sake the algorithm proposed by Zhdankin et al. (2013) for
the analysis of current sheets forming in MHD-turbulent plasmas, 
was extended to analyse the role and propertes of current sheets
formating in a much noisier kinetically turbulent plasma.
The applicability of this approach to the analysis of kinetically-turbulent 
plasmas is verified. Invesigated are, e.g., the effects of the choice of 
parameters on the current sheet recognition, viz. the threshold current 
density, the minimum current density and of the local regions around 
current density peaks.
The main current sheet properties are derived, their peak current 
density, the peak current carrier velocity (mainly electrons), the
thickness and length of the current sheets, i.e. also their aspect 
ratio (length/thickness). 
By varying the grid resolution of the simulations it is shown 
that, as long as the electron inertia is not taken into account,
the current sheets thin down well below ion inertial length
scale until numerical (grid-resolution based) dissipation 
stops any the further thinning.

\end{abstract}
\maketitle
\section{Introduction}
Dissipation of macroscopic energy into heat in the rarity of collisions in space and astrophysical plasmas is a major unsolved problem. In collisionless plasmas, the collisional dissipation scale is very small compared to the energy containing macroscopic scales, and thus macroscopic energy can not be dissipated into heat by collisions. Heating of collisionless plasmas is, nevertheless, quite common phenomena in space and astrophysics. For example, spacecraft observations of solar wind show the non-adiabatic dependence of its temperature on the distance from the surface of the sun. This phenomena indicates that a separate heating mechanism, which can not be accounted for merely by collisions, is operating inside the cluster of ionized particles departed from sun. Research carried out so far using theory, numerical computation and the satellite observations strongly suggest a crucial role of plasma turbulence
in the heating. Turbulence in collisionless plasmas transfer energy from macro to micro-scales associated with electrons and ions (their inertial length and gyro-radii) constituting the plasma. Collisionless plasma processes at the kinetic scales can, then, dissipate the energy.

At kinetic scales, a variety of kinetic plasma processes can play roles in the dynamics of turbulence. Numerical
simulations suggest that current sheets are formed at ion kinetic scales in plasma turbulence and co-exist with or are manifestation of kinetic plasma waves \cite{PhysRevX.9.031037}. Thicknesses of these current sheets range from ion to electron scales unlike the Kolmogorov scale current sheets formed in MHD simulations \cite{PhysRevLett.109.195001}. Spacecraft observations in solar wind and Earth's magnetosphere also reveal existence of current sheets in plasma turbulence \cite{Chasapis_2018}. Numerical simulations and spacecraft observations have further shown that a large part of the dissipation of the fluctuations and plasma heating occur in and around the current sheets \cite{PhysRevLett.109.195001,Chasapis_2018}. The kinetic plasma processes responsible for the dissipation and heating in the current sheets are, however, not well understood.

Several kinetic plasma processes in current sheets have been proposed to cause collisionless dissipation and heating in plasma turbulence. Magnetic reconnection in current sheets of plasma turbulence is one of the possible processes which can dissipate the turbulent fluctuations. It generates parallel electric fieldswhich can accelerate the charged particles \cite{Egdal}. Further, magnetic islands of varying size form and evolve in magnetic reconnection. Charge particles trapped inside contracting magnetic islands \cite{Drake} can be accelerated by Fermi-like processes. Stochastic ion heating \cite{2011ApJ...739...22M}, and Landau and cyclotron damping \cite{TenBarge_2013} in current sheets have also been considered for collisionless dissipation.

The kinetic plasma processes responsible for the dissipation in kinetic scale current sheets can be, directly or indirectly, influenced by plasma instabilities. For example, magnetic reconnection in current sheets formed in plasma turbulence is basically a tearing instability. Plasma instabilities in current sheets can generate large amplitude plasma waves which can lead to the stochastic ion heating \cite{Demchenko_1974}. Plasma instabilities can generate their own turbulence affecting the properties of the encompassing turbulence and providing anomalous dissipation.

Growth of plasma instabilities in current sheets and their nonlinear consequences for the dissipation depend upon free energy sources available in current sheets of turbulence as well as on physical parameters and structure of the current sheets. Free energy sources can arise either from spatial gradients of macroscopic variables (density, velocity and pressure) and/or by non-Maxwellian velocity distribution of the plasma particles.
Therefore current sheets forming in plasma turbulence need to be characterized in terms of free energy sources, their gradient scale lengths and/or non-Maxwellean features, and physical parameters and structure of current sheets to understand the role of plasma instabilities in collisionless dissipation.

The characterization of current sheets first requires identification of individual current sheets embedded in a turbulent background and then estimations of their characteristics. Manual identification and characterization of current sheets in turbulence are tedious processes and prone to human errors. This makes the case for the development of computer programs which can speed-up and automate the identification and characterization of individual current sheets embedded in a turbulent background.

Zhdankin et al. \cite{Zhdankin_2013} suggested an algorithm to identify and characterize current sheets in magnetohydrodynamic (MHD) turbulence. The algorithm implemented in the program requires to choose optimum values of three parameters for a statistically meaningful identification and characterization of current sheets. These parameters are: (1) threshold current density, (2) size of the local region surrounding an individual current sheet and (3) minimum (or boundary) value of the current density in a current sheet. The values of these algorithm parameters can not be prescribed universally and need to be chosen on the case-basis guided by the dependence of the results on the algorithm parameters.

We programmed in PYTHON a computer code similar to that Zhdankin et al. \cite{Zhdankin_2013}  but additionally 
allowing to parametrize the specifics of the kinetic plasma turbulence. 
The new parameter is an alternative to the current density threshold parameter and is useful to cross-check the results obtained by using the current density threshold. Although its value is case specific, it can be better estimated from the turbulence data. 

In this work, we study the dependence of the identification and characterization of current sheets on the algorithm parameters by applying a newly developed PYTHON-based computer program to the turbulence data generated by 2-D PIC-hybrid simulations. Then, we characterize the current sheets formed in the PIC-hybrid simulations.


The paper is organized as follow. In Section \RN{2}, we discuss the 2-D PIC-hybrid simulations of collisionless plasma turbulence used to generate the turbulence data for the analysis of current sheets. Section \RN{3} discusses the effect of varying algorithm  parameters  on the current sheet identification and characterization. A new algorithm parameter is also introduced in section \RN{3}. The results of current sheet characterization are presented in section \RN{4}. Section \RN{5} presents the discussion. The paper is concluded in section \RN{6}. An overview of Zhdankin's algorithm of current sheet identification and characterization is given in the appendix. 

\section{2-D PIC-hybrid simulations of collisionless plasma turbulence}
We characterize current sheets self-consistently formed in 2-D PIC hybrid simulations of collisionless plasma turbulence carried out by Jain et al. \cite{jain2020}.
The simulations are initialized with random phased fluctuations of magnetic field and plasma velocity imposed on a uniform and isotropic background plasma in an x-y plane. A uniform magnetic field  $B_0\hat{z}$ perpendicular to the simulation plane is
applied.
\\



The root mean square value of the fluctuations is $B_{rms}/B_0 = 0.24$ 
and are initialized in the wave number range 
$|k_{x,y}d_i| < 0.2 (k_{x,y} \ne 0)$. 
Electron and ion plasma beta are $\beta_e = 0.5$ and $\beta_i = 0.5$, respectively. 
The simulation box size for is $256d_i \times 256d_i$ with $512 \times 512$ grid points, 500 particles per cell and the time step $\Delta t = 0.01\Omega^{-1}_{ci}$. 
Simulations are also carried out by varying number of  grid points, particles per cell and time step to check the effect of numerical parameters. 
Boundary conditions are periodic in all directions. 

\subsection{Data selection}
\begin{figure}[hbt!]
\centering
\includegraphics[width=7.7cm,height=5.7cm]{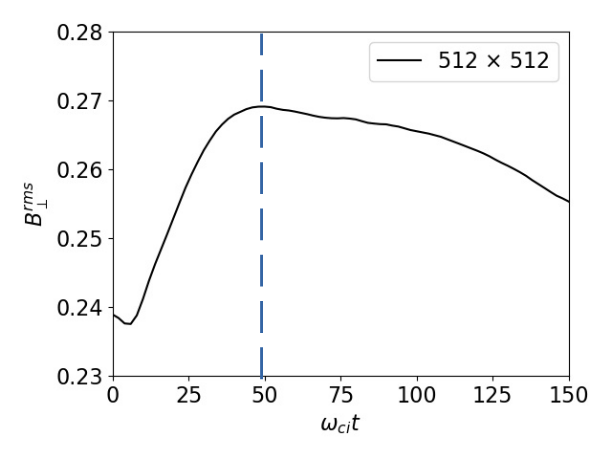}
\vspace{-0.5cm}
\caption{Evolution of $B_{\perp}^{rms}$ in the hybrid simulations.
A dashed vertical line is drawn at  $\omega_{ci}t = 50$.  \label{fig:bperp_rms}}
\end{figure}

\begin{figure}[hbt!]
\centering
\includegraphics[width=8cm,height=5.5cm]{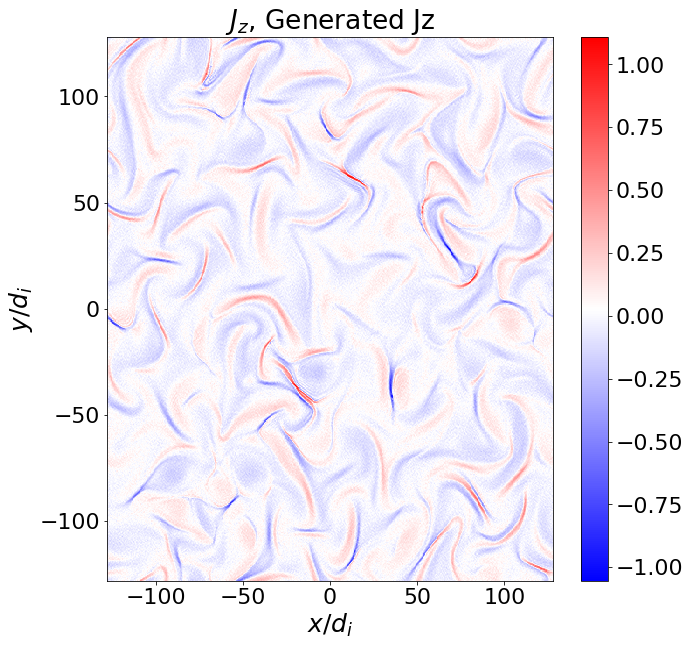}
\caption{Current density $J_z$ in $x-y$ plane at $\omega_{ci}t = 50$.  \label{fig:jz_t50}}
\end{figure}

Figure \ref{fig:bperp_rms} shows that the rms value of perpendicular (to the external magnetic field in the z-direction) magnetic field peaks at $\omega_{ci}t= 50$.  Since current sheets store magnetic energy, the peak in the perpendicular RMS magnetic field can be taken as an indicator of peak activity of current sheet formation. Parallel current density at $\omega_{ci}t =50$, shown in Fig. \ref{fig:jz_t50}, displays  presence of current sheets in plasma turbulence. Later these  current sheets gets signiﬁcantly disrupted \cite{jain2020}. As our objective in this research work is identification and characterization of current sheets before they get disrupted, we choose $\omega_{ci}t = 50$ as a time at which we characterize current sheets. 

\section{Choosing algorithm parameters}
In order to identify and characterize current sheets in the plasma turbulence we use Zhdankin et al. (2013) algorithm \cite{Zhdankin_2013} which is divided in two parts, identification and characterization (see details in appendix). The algorithm identifies current sheets by detecting local maxima of current density above a threshold value $J_{thr}$ in a local region containing $(2n+1)^2$ points (see Fig. \ref{fig:local_region} in the appendix for the definition of $n$). Then it finds points collectively connected to the local maxima and where current density is above a minimum (boundary) value $J_{min}$. Therefore  the algorithm has three free parameters to choose, viz., $J_{thr}$, $n$ and $J_{min}$.

Threshold current density is required to separate out the current sheets from background turbulence and can be expected to be several times the root-mean-square (RMS) current density. At the same time, the simulations reveal that the ratio $|u_{ez}|/u_{iz}^{rms}$ is much larger than unity only in the kinetic scale current sheets \cite{jain2020}. For the simulations used here, $|u_{ez}|/u_{iz}^{rms} \gtrsim 8$ \cite{jain2020}. Therefore threshold on the value of $|u_{ez}|/u_{iz}^{rms}$ can also be used  as a criterion to isolate current sheets from the background turbulence. We have implemented both methods of threshold in our PYTHON program.

\begin{figure}[!ht]
\minipage{0.31\textwidth}
\includegraphics[width=6cm,height=4.5cm]{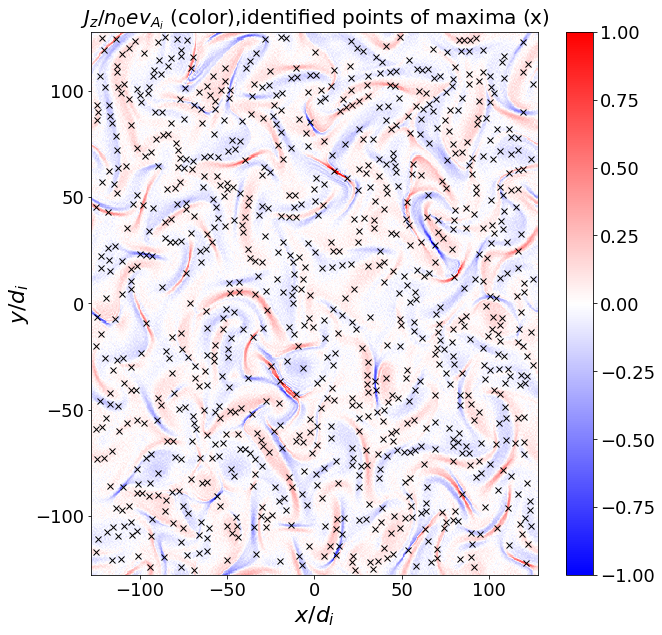}
\vspace{-7mm}
  \caption*{$J_{thr}=J_{rms},n=5$ }
\endminipage\hfill
\minipage{0.31\textwidth}
\includegraphics[width=6cm,height=4.5cm]{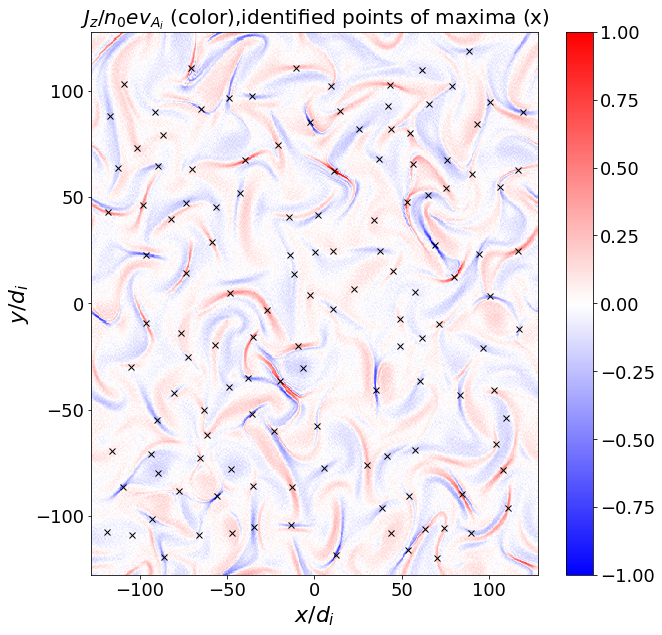}
\vspace{-7mm}

  \caption*{$J_{thr}=J_{rms},n=15$ }
\endminipage\hfill
\minipage{0.31\textwidth}
\includegraphics[width=6cm,height=4.5cm]{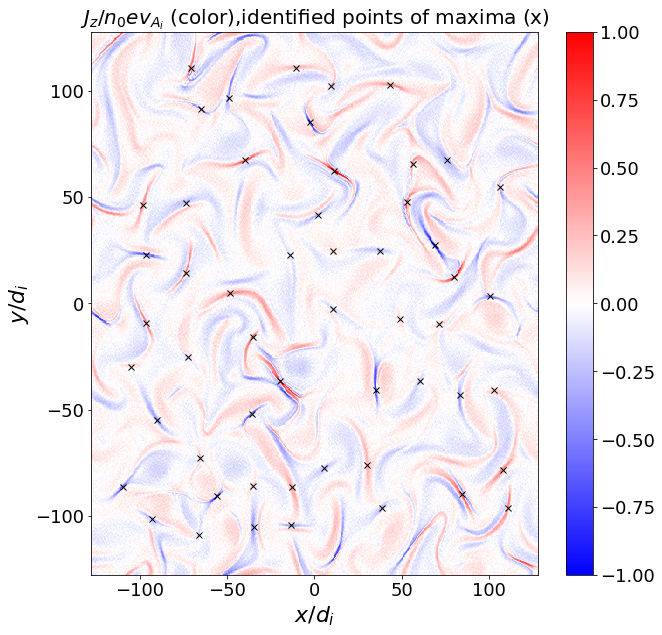}
\vspace{-7mm}

  \caption*{$J_{thr}=J_{rms},n=25 $}
\endminipage\
\vspace{5mm}

\minipage{0.31\textwidth}
\includegraphics[width=6cm,height=4.5cm]{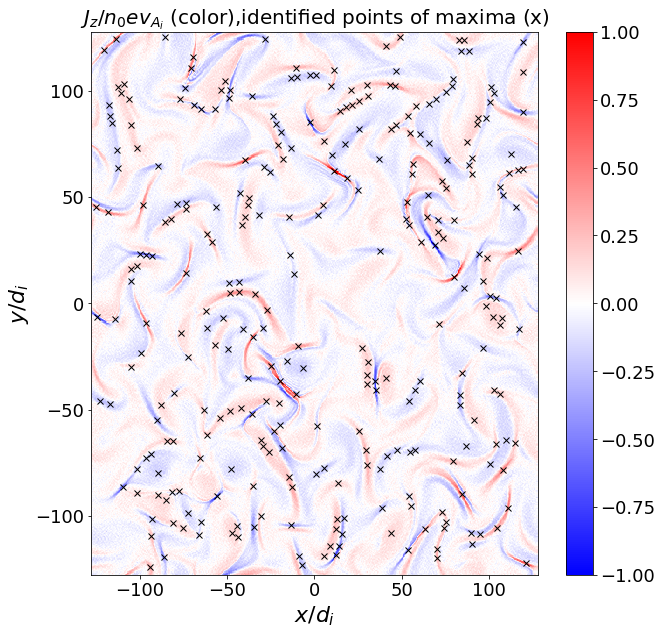}
\vspace{-7mm}

  \caption*{$J_{thr}=2J_{rms},n=5 $}
\endminipage\hfill
\minipage{0.31\textwidth}
\includegraphics[width=6cm,height=4.5cm]{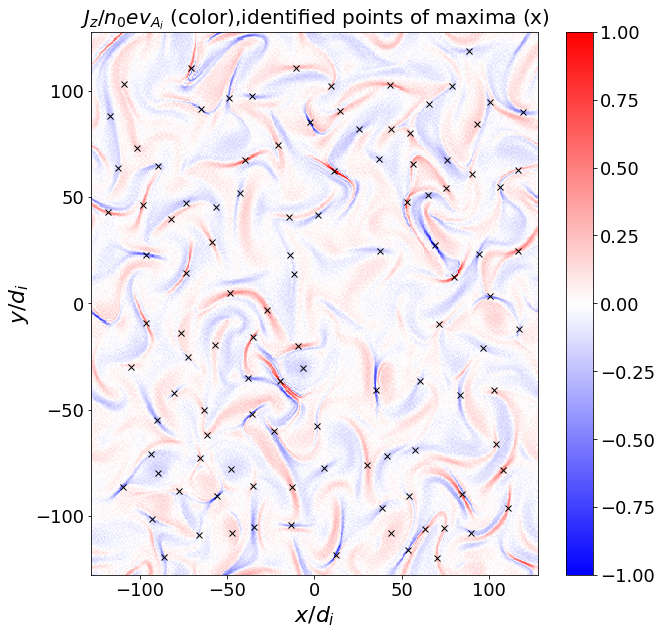}
\vspace{-7mm}

  \caption*{$J_{thr}=2J_{rms},n=15$ }
\endminipage\hfill
\minipage{0.31\textwidth}
\includegraphics[width=6cm,height=4.5cm]{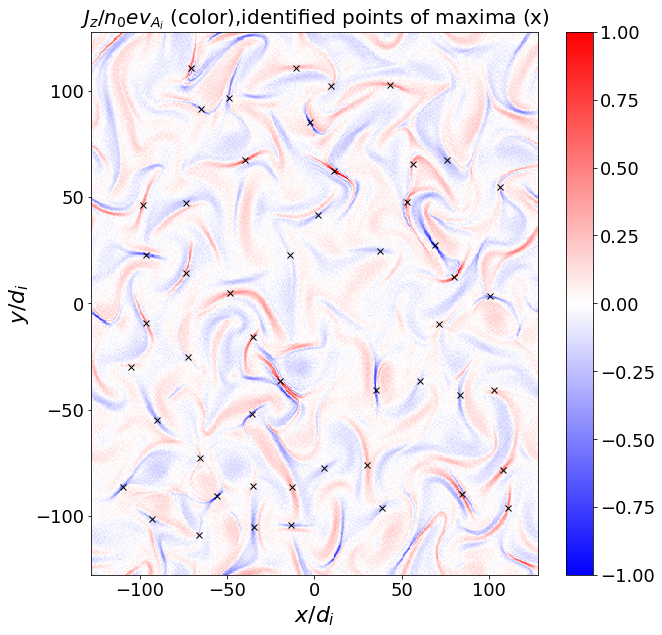}
\vspace{-7mm}
 \caption*{$J_{thr}=2J_{rms},n=25$ }
\endminipage\
  \caption{Parallel current density $J_z$ (color) in the simulation plane at $\omega_{ci}t=50$. Locations of local maxima ('$\times$') found by the algorithm are over plotted for $J_{thr}=J_{rms}$ (top row) and $2 J_{rms}$ (bottom row). In a row, value of n increases, $n=5, 15, 25$ (left to right) \label{fig:local_maxima}.
}
\end{figure}

Two of the algorithm parameters, the size of local region surrounding local maxima $n$ and  threshold current density $J_{thr}$ (or $u_{ez}/u_{iz,rms}$) are required for finding the locations of local maxima in current density. Therefore we vary these two parameters and plot  the locations of local maxima over current density in Fig. \ref{fig:local_maxima}. 
As it can be seen from  Fig. \ref{fig:local_maxima} that for  $J_{thr}=J_{rms}$ and $n=5$ , algorithm found unreasonable number of local maxima. The reason is that for small values of  threshold, the algorithm does not filter out fluctuations very well and many local maxima associated with these fluctuations are still present in the filtered data. Although peak current densities associated with the fluctuations are typically smaller than those associated with current sheets, they are not discarded by the algorithm as small values of $n$ may not allow their comparison with the current sheet associated peaks. For larger value of threshold current density $J_{thr}=2J_{rms}$ unwanted fluctuation peaks are filtered better and the impact is visually tangible for $n=5$ in  Fig. \ref{fig:local_maxima}. Also by keeping $J_{thr}= J_{rms}$  and increasing the value of $n$, the number of unwanted fluctuation peaks are reduced as many of them are now discarded in comparison to current sheet peaks. For values of $J_{thr} =J_{rms}$ and $n=25$, not all but a reasonable number of peaks, sufficient for statistical analysis, are detected and results are almost the same for $J_{thr}=J_{rms}$ and $2 J_{rms}$ when $n=25$.

\begin{figure}[!ht]
\vspace{-1mm}
\minipage{0.31\textwidth}
\includegraphics[width=6cm,height=4.5cm]{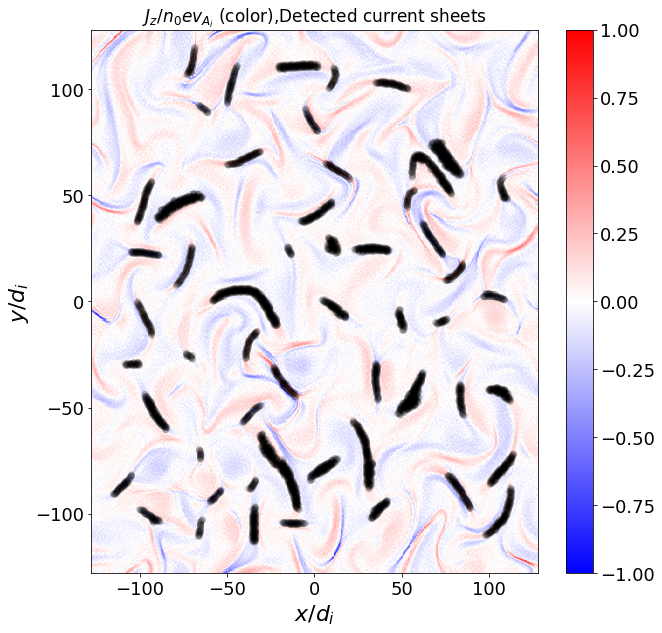}
\vspace{-0.4cm}

\endminipage\hfill
\minipage{0.31\textwidth}
\includegraphics[width=6cm,height=4.5cm]{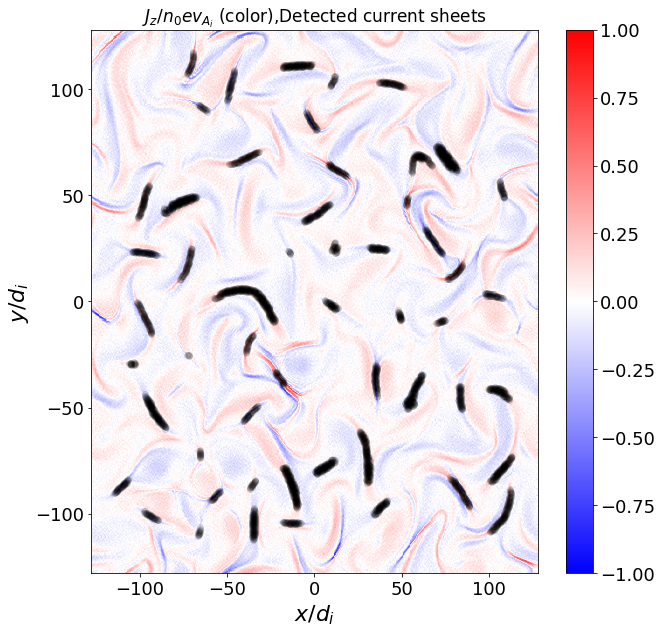}
\vspace{-7mm}
\endminipage\hfill
\minipage{0.31\textwidth}
\includegraphics[width=6cm,height=4.5cm]{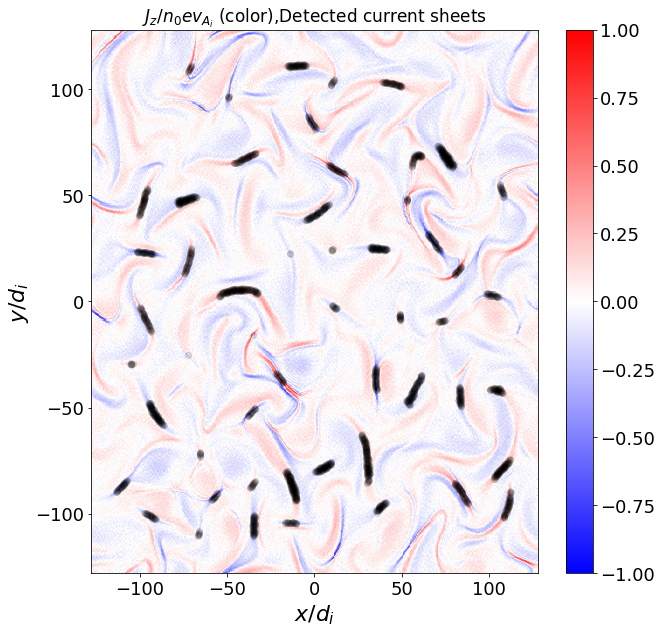}
\vspace{-7mm}

\endminipage\

  \caption{Detected current sheet points (black dots) for  $J_{thr} = J_{rms}$, $n=25$ and $J_{min,i}/J_{max,i}=(0.4, 0.5, 0.6)$ (left to right) plotted over $J_z$ (color).
\label{fig:detected_cs}}
\end{figure}

\begin{figure}[!hbt]
\minipage{0.31\textwidth}
\includegraphics[width=6cm,height=4.5cm]{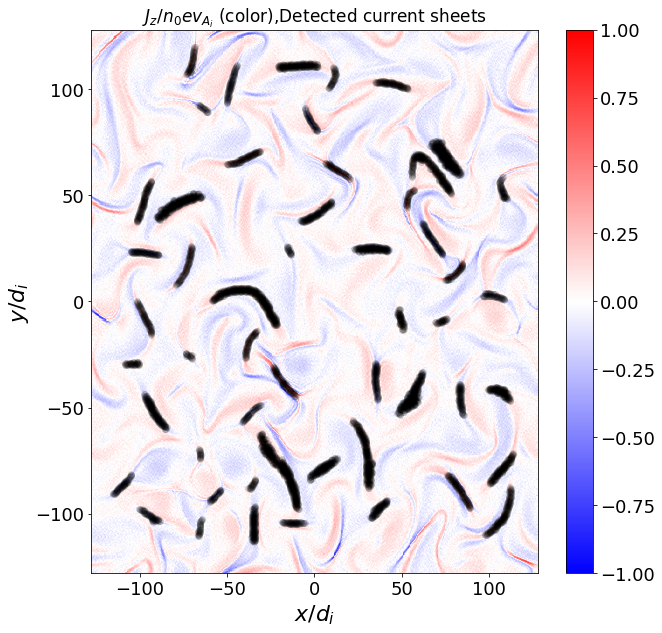}
\endminipage\hfill
\minipage{0.31\textwidth}
\includegraphics[width=6cm,height=4.5cm]{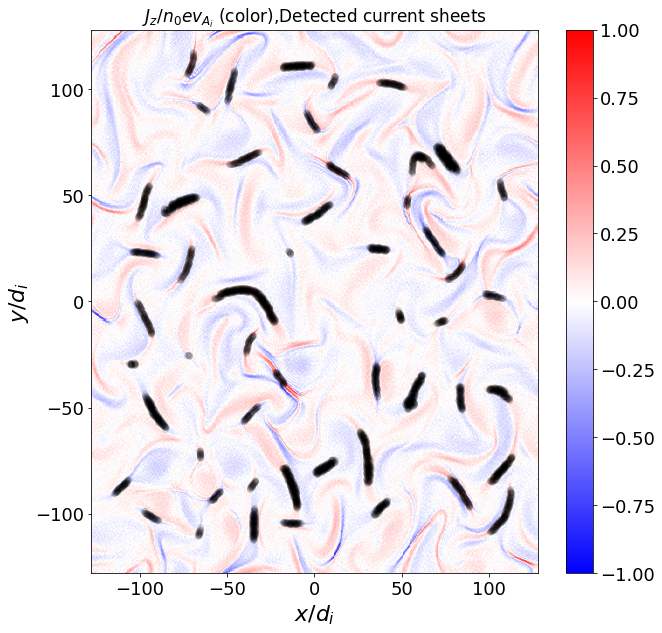}
\endminipage\hfill
\minipage{0.31\textwidth}
\includegraphics[width=6cm,height=4.5cm]{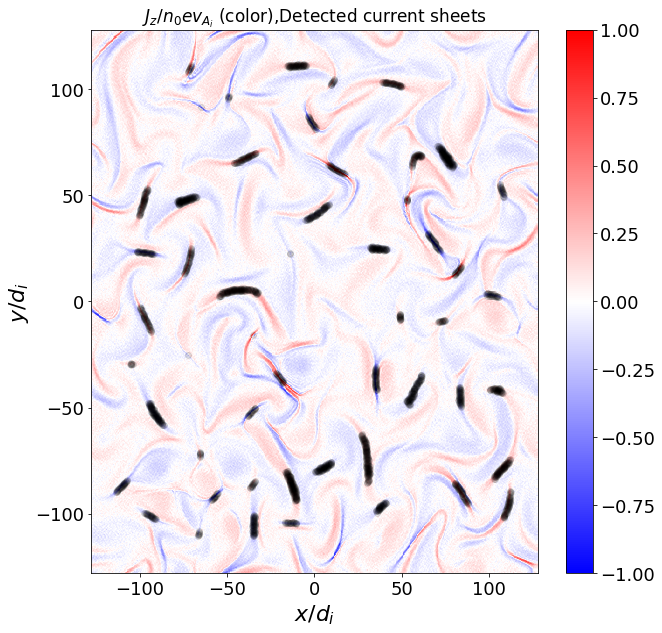}
\endminipage\

\caption{Detected current sheet points (black dots) plotted over $J_z$ (color) using threshold on the ratio $u_{ez}/u_{iz,rms}$ with a threshold value $[u_{ez}/u_{iz,rms}]_{thr}=7.25$. Other parameters are $n=25$ and $J_{min,i}/J_{max,i}=(0.4, 0.5, 0.6)$ (left to right).
\label{fig:detected_cs1}}
\end{figure}

After the detection of peak location of current density  for each current sheet, algorithm finds  points which are collectively connected to the peak and have magnitude $J_{z}$ above the $J_{min}$ value (minimum value of current density).
Here we choose value of $n=25$ for the size of local region ,  $J_{thr}=J_{rms}$ for  the thresholds and $J_{min,i} = (0.4,0.5,0.6) J_{max,i}$ ('i' is the current sheet index). Fig \ref{fig:detected_cs} shows the effect of changing $J_{min,i}$ on the identification of points belonging to each current sheet. Also it can be seen that by increasing the $J_{min,i}$ value, the number of points belonging to each current sheet  are decreased so as to reduce current sheet lengths. We also show in Fig. \ref{fig:detected_cs1} the results of current sheet detection using a threshold value of the ratio $[u_{ez}/u_{iz,rms}]_{thr}=7.25$. It is clear that the results obtained by using the two threshold methods are almost identical. 
Fig. \ref{fig:detected_cs} and \ref{fig:detected_cs1} show that appropriate values of  $J_{min,i}/J_{max,i}$ for current sheets characterization are 0.4 and 0.5 and the results for these values should be compared. We, however, choose $J_{min,i}/J_{max,i}=0.5 $ in this research work and defer the current sheet characterization for  $J_{min,i}/J_{max,i}=0.4$ to future work.

\onecolumngrid
\onecolumngrid

{}

\newpage
\section{Characterization of current sheets}

\begin{figure}[!hbt]

\includegraphics[width=7.4cm,height=5.5cm]{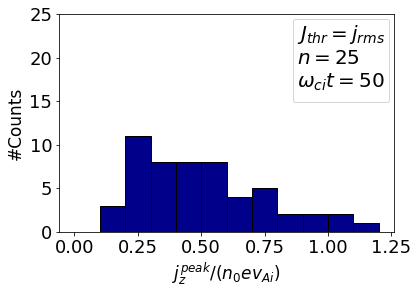}
\includegraphics[width=7.4cm,height=5.5cm]{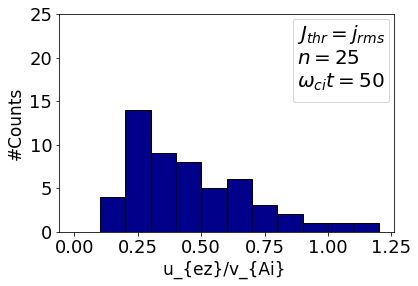}

 \caption{Distribution of peak current density (top) and peak parallel electron velocity (bottom) in current sheets at $\omega_{ci}t=50$ for $J_{thr} = j_{rms}$, $n=25$ and $J_{min,i}=0.5 J_{max,i}$. \label{fig:hist_jzpeak}}
\end{figure}

Fig. \ref{fig:hist_jzpeak} shows distribution of peak current density $J_z^{peak}$ and the associated peak parallel electron velocity ($u_{ez}^{peak}=u_{iz}^{peak}-J_{z}^{peak}/\rho^{peak}$) of current sheets for the algorithm parameters  $J_{thr}=J_{rms}$, $n=25$ and $J_{min,i}/J_{max,i}=0.5$.  Majority of current sheets have peak current density in the range 0.2-0.4 $n_0ev_{Ai}$ with maximum peak current density reaching up to 1.1 $n_0ev_{Ai}$. Peak parallel  electron bulk velocity also has a similar distribution  with most values in the range 0.2 to $0.4 v_{Ai}$ and maximum reaching 1.1 $v_{Ai}$. The similarity in the distributions of peak current density and peak parallel electron bulk velocity arises from the fact that ion bulk velocity is much smaller than electron bulk velocity and density normalized to background density is close to unity in current sheets. Therefore current in the sheets is almost entirely due to the electron bulk velocity.

\begin{figure}[!ht]
  \includegraphics[width=7.2cm,height=5.5cm]{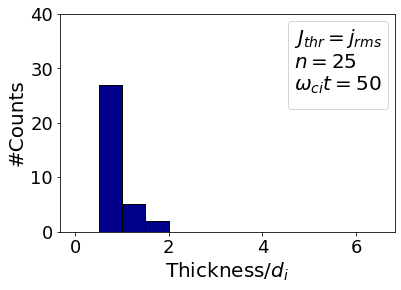}

  \includegraphics[width=7.6cm,height=5.5cm]{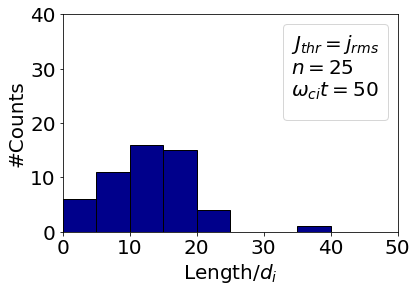}

  \includegraphics[width=7.2cm,height=5.5cm]{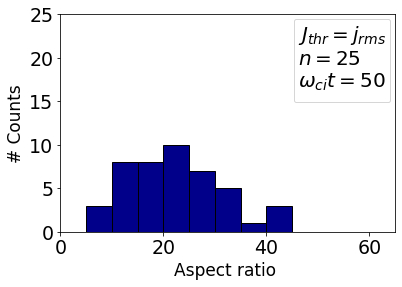}
\caption{Distribution of half thickness (top), length (middle) and aspect ratio (bottom) of current sheets at $\omega_{ci}t=50$ for $J_{thr} = j_{rms}$, $n=25$ and $J_{min,i}=0.5 J_{max,i}$. \label{fig:hist_thickness}}
\end{figure}

\begin{figure}[!ht]
\includegraphics[width=7.2cm,height=5.08cm]{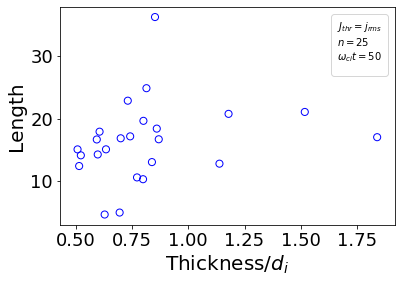}
\caption{Length vs. thickness of characterized current sheets. \label{fig:length_vs_thickness}}
\end{figure}

Fig. \ref{fig:hist_thickness} show distributions of half-thickness, length and  aspect ratio (length/half-thickness) of current sheets for $n=25$, $J_{thr}=J_{rms}$ and $J_{min,i}=0.5\,J_{max,i}$. Majority of current sheets have thicknesses close to  0.5 ion inertial length which is grid spacing for the simulation data. Characterization of current sheets formed in PIC-hybrid simulations with higher grid resolutions shows that the peak in the distribution of half-thickness is always near the grid resolution of the simulations. Therefore, current sheets formed in PIC-hybrid simulations of kinetic plasma turbulence have the tendency to thin down to below ion inertial length until the thinning is stopped by the numerical effects at the grid scales. The length of the majority of current sheets, on the other hand, lies in the range of $5\,d_{i}$ to $25\,d_{i}$, mostly around $15\,d_{i}$. Few current sheets have lengths reaching upto  $40\,d_{i}$. Fig. \ref{fig:length_vs_thickness} shows that the length of the current sheets in the range 10-25 $d_i$ seems to have a positive correlated with the half-thickness. It means thicker current sheets are lengthier.   

For spatial gradient driven instabilities, aspect ratio, the ratio of the length and half-thickness, of current sheets is an important parameter.  Distribution of aspect ratio in Fig. \ref{fig:hist_thickness} shows that the values of aspect ratio for majority of current sheets are concentrated in the range of 10 to 40, larger than $2\,\pi$ typically required for many spatial gradient driven instabilities. 

\section{Discussion}

We characterized current sheets in terms of their peak current density, peak parallel electron velocity, thickness, length and aspect ratio in two dimensional
kinetic plasma turbulence in which variations along the direction of the large scale magnetic field, and thus also along the direction of the current
in the sheets, were neglected. The characterization results can be physically interpreted within the limitations of the two-dimensionality and lack of electron inertia in the hybrid plasma model used here.


Our results of current sheet characterization show that current sheets continue to thin down below ion inertial length as much as allowed by the
grid resolution of the hybrid simulations and then get distorted by artificial numerical effects. It means current sheets prefer to thin down below ion inertial length rather than develop 2-D ion-scale plasma instabilities with perpendicular wave vectors. This is in contrast with other hybrid simulations of collisionless plasma turbulence in which current sheets formed in the turbulence do not thin down below ion inertial length but develop perpendicular tearing instability leading to magnetic reconnection when their thicknesses are close to the ion inertial length (much larger than grid scale) \cite{Franci_2017,Papini_2019}. This difference is due to the values of plasma resistivity used in the hybrid simulations. In other hybrid simulations \cite{Franci_2017,Papini_2019},
value of the plasma resistivity is fine tuned to set the collisional dissipation scale length close to an ion inertial length causing development of tearing instability when current sheets thin down to ion inertial length. On the other hand, in the hybrid simulations used in this paper, plasma resistivity has been chosen to be zero (with a very small time step to stabilize the wave modes at the grid scales) not allowing the growth of the tearing instability. Instead, current sheets continue to thin down to below ion inertial length.

The use of plasma resitivity in hybrid simulations is for the numerical purposes and thus is artificial. Owing to much smaller plasma resistivity in most physical system, current sheets therein are likely to thin down below ion inertial length (if not hindered by 3-D ion scale plasma instabilities) rather than developing  2-D tearing instabilities. Continuous thinning of current sheets below ion inertial length will ultimately lead to current sheet thicknesses of the order  electron scale lengths where the thinning can be
stopped by physical effects, for example, by finite electron inertia, rather than artificial resistivity. Vlasov-hybrid simulations with reduced ion to electron mass ratio shows that current sheets indeed thin down to few electron inertial length \cite{califano2020}.

Hybrid simulations of collisionless plasma turbulence  show the development
of electron shear flow as one of the free energy sources in current sheets \cite{jain2020}. If current sheets thin down to electron inertial length, they may become unstable to electron inertia driven electron shear flow instabilities (ESFI) \cite{chap5a}.
ESFI grows as a tearing
instability for weak guide magnetic field (parallel to the direction of electron flow/current) and/or current
sheet thickness less than or of the order of an electron inertial length \cite{chap5a}. It grows as
a non-tearing instability for strong guide field and/or thicker current sheets \cite{chap5a}. Since current sheets in collisionless plasma turbulence thins down from thicknesses much larger than an electron inertial length, they are susceptible to non-tearing ESFI as long as thicknesses do not reduce to an
electron inertial length. Non-tearing ESFI, however, may not grow if the thinning process is faster than the
instabilities. In that case, current sheets are expected to thin down to electron inertial length or below and
the tearing ESFI can grow leading to magnetic reconnection at electron scales.

For the tearing ESFI to grow in a current sheet of half-thickness $L_{cs}$, the unstable wavelength $\lambda$ along the
current sheet length must satisfy the instability condition $\lambda/L_{cs}>2\pi$ \cite{chap5b}. For $\lambda \sim l_{cs}$, where $l_{cs}$ is the length of current sheet, we get a condition on the aspect ratio of the current sheet as $l_{cs}/L_{cs} > 2\,\pi$.
Results of the current sheet characterization show that majority of the current sheets have the
aspect ratio $\sim$ 20 even when their thicknesses are of the order of ion inertial length. Thinner current sheets,
as expected in collisionless plasma turbulence, would have much larger asppect ratios and can therefore
become unstable to the tearing ESFI if thickness reaches an electron inertial length.
 More simulations of collisionless plasma turbulence employing plasma models which include the full electron scale physics (for example, hybrid plasma model with electron inertia implemented in the code CHIEF \cite{munoz2018} or the fully kinetic model for electrons) and characterization of current sheets formed therein are required to correctly determine the properties of the current sheets.

Theoretical estimates in the limit of unmagnetized ions supporting the results of the PIC-hybrid simulations of collisionless plasma turbulence give a scaling relation $|u_{ez}/u_{iz}| \sim |1-d_i^2/L_{cs}^2|$ \cite{jain2020}. For electron scale current sheets, $L_{cs} \sim d_e$, $|u_{ez}/u_{iz}| \sim m_i^2/m_e^2 >> 1$ ($m_i$ and $m_e$ are ion and electron masses respectively), which brings in relative streaming of electrons and ions as another free energy source in current sheets. Electron-ion streaming driven instabilities are current aligned and therefore 3-D simulations of collisionless plasma turbulence are required to pin-point their role in the turbulence.  




 \section{Conclusions}
We  identified and characterized current sheets forming in a kinetic plasma turbulence 
by applying the algorithm of Zhdankin et al. (2013)~\cite{Zhdankin_2013} (originally used to  
diagnose current sheets in MHD-turbulent plasmas)  to the
results of hybrid-kinetic code simulations of collisionless plasmas. 
The algorithm parameters, viz., threshold value $J_{thr}$, the size of local region surrounding local maxima $n$ and 
the minimum current density $J_{min,i}$ were determined by experimentation on 
PIC-hybrid simulation of plasma. 
We characterized the current sheets in terms of peak current density, thickness, length and aspect
ratio and the results provide insights in the nature of current sheets in collisionless plasma turbulence. 
Considering the limitation of simulation, two dimensionality and neglect of electron scale physics, statistical analysis shows that current sheets tend to thin down to grid scale. This thinning process can leads to two possible plasma instabilities which are electron shear flow instabilities and electron-ion streaming instabilities. However, to understand the exact fate of current sheet's thinning process the three dimensional plasma simulation which include electron scale physics should be carried out.

\section*{Appendix: Algorithm}
\subsection{Identification of current sheets}
\begin{figure}[hbt!]
\centering
\includegraphics[width=7.4cm,height=5.5cm]{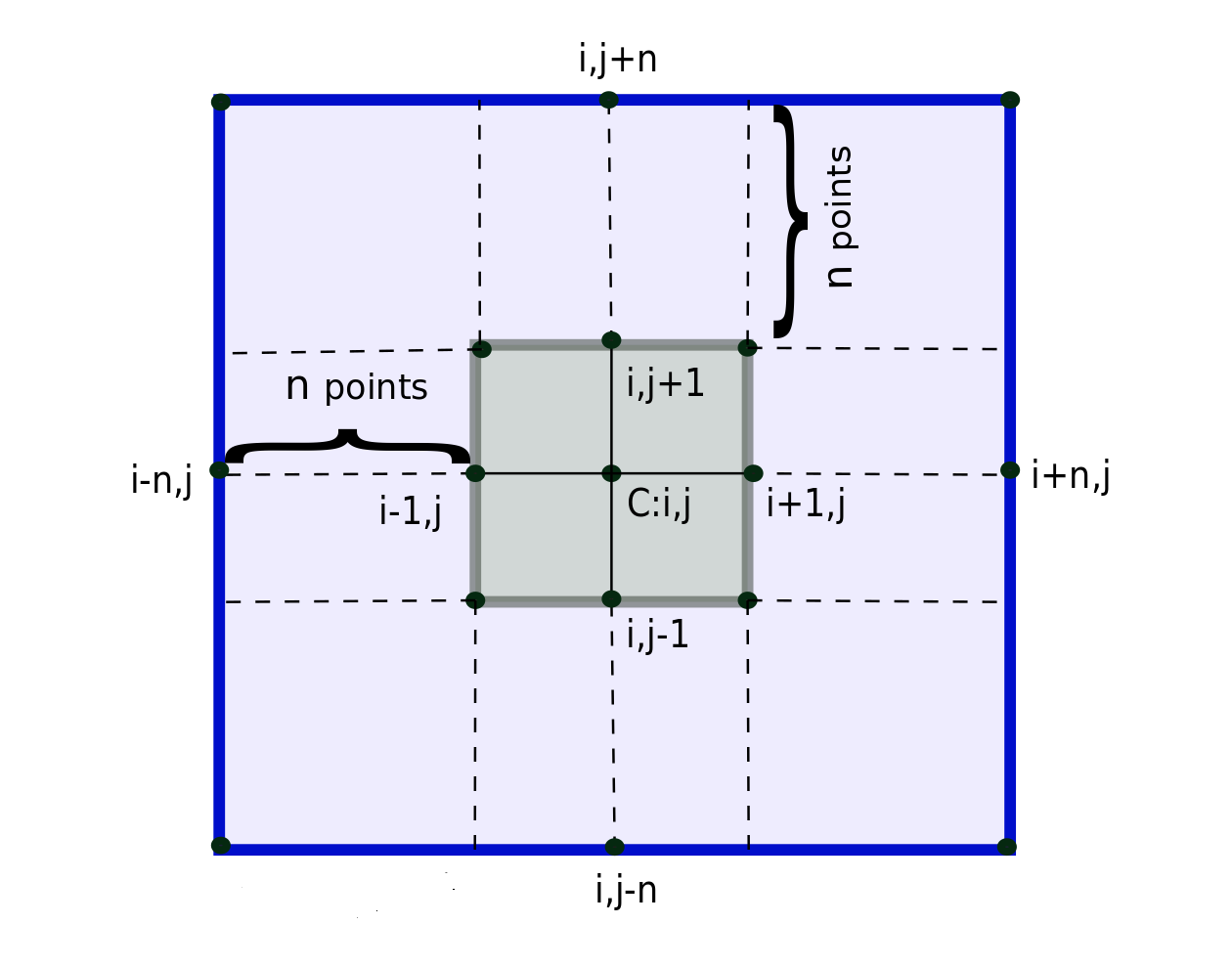}
\caption{A local region surrounding a candidate point $C(i,j)$ and containing $(2n+1)^2$ points($n$ points on either side of point $C$) in $2-D$.  In Zhdankin's algorithm\cite{Zhdankin_2013}, magnitude of current density at candidate point $C(i,j)$ is checked for it to be local maximum over this local region. \label{fig:local_region}  }
\end{figure}

 A starting step in the identification of current sheets is to look for local maxima in current density magnitude. But current density in plasma turbulence can have several tiny peak(due to the turbulent fluctuations) which  don't qualify as  current sheets. This is achieved in Zhdankin et al.(2013)-algorithm \cite{Zhdankin_2013} by defining a threshold current density $J_{thr}$  sufficiently larger than the typical fluctuation level and selecting only the data points for which magnitude of current density is greater than the threshold value.  Then we search for local maxima only at the selected points.  Hence, algorithm scans through all the data points where current density is above  the specified $J_{th}$ and selects those points which are local maxima in the region surrounded by $n$ points on either side of the candidate point in each direction.  Fig. \ref{fig:local_region} shows a candidate point indexed by $(i,j)$ surrounded by $(2n+1)^2$ points of local region in the 2-D case.

\begin{figure}[!hbt]
\includegraphics[width=16.4cm,height=5cm]{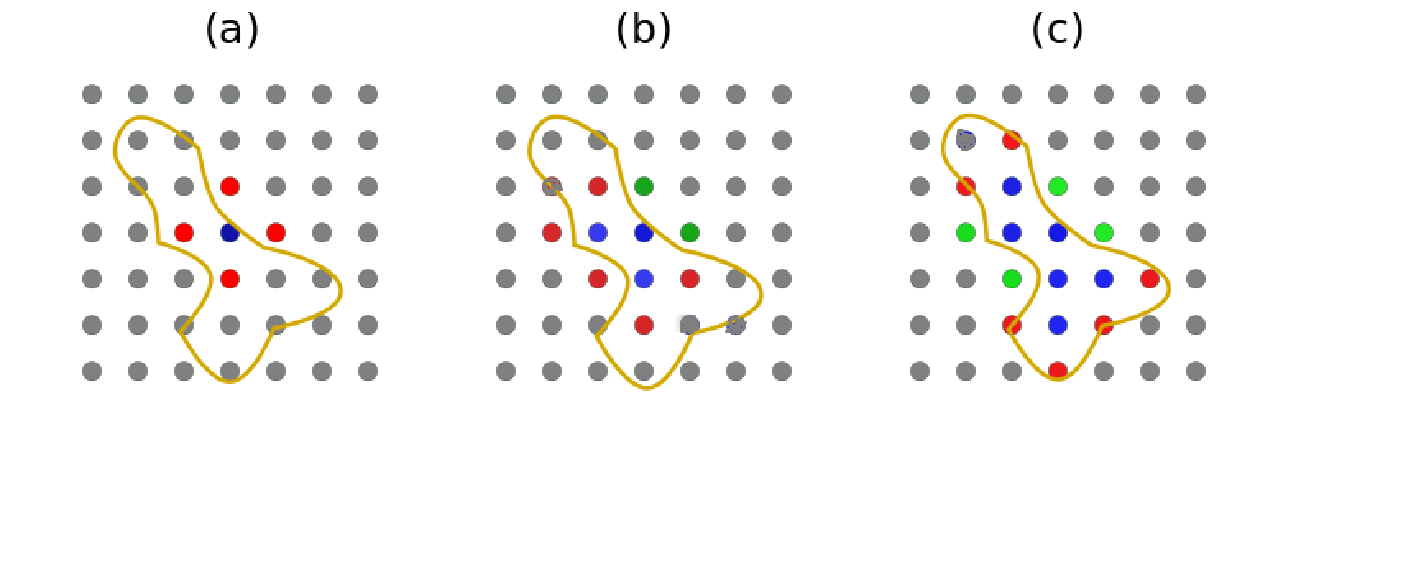}
\vspace{-1.3cm}
\caption{ A series of steps  in Zhdankin's algorithm \cite{Zhdankin_2013}, from (a) to (c), to find points belonging to the i-th current sheet (region enclosed by closed contour ) where the condition $J>J_{min,i}=J_{max,i}/2$ is satisfied. The color of a grid point (filled circles) at a given step indicates if the condition is satisfied (blue), not satisfied (green), will be checked (red) or will not be checked (gray) at that grid point. 
The blue point in (a) is a local maximum in current density from where the procedure starts. \label{fig:cs_points_procedure}   }
\end{figure}

In a smooth current sheet the current density drops from its peak value to vanishingly small values in distances of the order of half-thickness of the current sheet. Therefore points belonging to the current sheet can be defined to be the points where the current density is larger than a sufficiently small value and which are collectively connected to the point of peak current density (see Fig. \ref{fig:cs_points_procedure}). The condition that is defined in  algorithm  for identifying current sheet points is that each point has the magnitude of the current density greater than a minimal value of $J_{min,i} = J_{max,i}/2$ (i denotes the ith local maxima or current sheet).

\subsection{Characterization of current sheets}

For calculating  current sheet thickness first the direction of the most rapid descent of the current density 
from the peak value is obtained by calculating the Hessian matrix values of parallel current 
density and its eigenvector at the peak. 
The direction of the eigenvector associated with the largest magnitude eigenvalue of the Hessian 
determines the direction of the most rapid descent.
Then the distance along this direction  from the current sheet peak to the point where current density 
drops to a value of $J_{min,i}$ is calculated. 
The same procedure is repeated in the opposite direction.
Both two distances together are taken as the total thickness of the current sheet. 
In order to obtain the length of the current sheet the longest distance between any pair of two points 
of the current sheet is determined by iterating over all  points in the $xy$ cross section of the 
current sheet. 
This method is accurate unless the $xy$ cross section is strongly curved.  
The second eigenvector of the Hessian matrix used to calculate the thickness of each 
current sheet, can also be used  to calculate the length. 
This would be, however, less accurate since in the case of curved current sheets 
the sheet boundary might quickly be reached.

\onecolumngrid

\begin{acknowledgments}
This work was supported by the German Science Foundation, DFG, project JA 2680 / 2-1.
\end{acknowledgments}

\bibliography{references}

\end{document}